\documentclass[aps,prl,twocolumn,amsmath,amssymb,floatfix,superscriptaddress]{revtex4-2}
\usepackage{graphicx}
\usepackage{bm}
\usepackage{hyperref}
\usepackage{braket,bbold,color}
\usepackage{tcolorbox}

\newcommand{\bs}[1]{\boldsymbol{#1}}
\begin{document}

\title{Dislocation Non-Hermitian Skin Effect}

\author{Frank Schindler}
\affiliation{Princeton Center for Theoretical Science, Princeton University, Princeton, NJ 08544, USA}

\author{Abhinav Prem}
\affiliation{Princeton Center for Theoretical Science, Princeton University, Princeton, NJ 08544, USA}

\begin{abstract}
We demonstrate that crystal defects can act as a probe of intrinsic non-Hermitian topology. In particular, in point-gapped systems with periodic boundary conditions, a pair of dislocations may induce a non-Hermitian skin effect, where an extensive number of Hamiltonian eigenstates localize at only one of the two dislocations. An example of such a phase are two-dimensional systems exhibiting weak non-Hermitian topology, which are adiabatically related to a decoupled stack of Hatano-Nelson chains. Moreover, we show that strong two-dimensional point-gap topology may also result in a dislocation response, even when there is no skin effect present with open boundary conditions. For both cases, we directly relate their bulk topology to a stable dislocation non-Hermitian skin effect. Finally, and in stark contrast to the Hermitian case, we find that gapless non-Hermitian systems hosting bulk exceptional points also give rise to a well-localized dislocation response.
\end{abstract}

\maketitle

\emph{Introduction}--- 
The ten-fold way~\cite{Kitaev_2009,Ryu_2010,Chiu:2016aa} enumerates all possible $d$-dimensional ``strong" topological phases protected by the ten Altland-Zirnbauer symmetry classes~\cite{AZclass}. Strong phases are characterized by a quantized topological invariant and protected gapless surface states which are stable against local symmetry-preserving perturbations and disorder~\cite{HasanKaneColloq,qizhangrmp,hasanmoorerev}. Besides strong phases, there exist ``weak" topological indices~\cite{Kane07,Roy2009,Moore07,noguchi2019} which derive from invariants defined on submanifolds of the Brillouin Zone (BZ) and are sensitive to disorder since they require lattice translation symmetry. These phases, which can be adiabatically connected to stacks of lower-dimensional topological phases, nevertheless display robust topological features~\cite{Mong2012,AdyWeak,morimoto2014}, including gapless edge modes along symmetry-preserving boundaries. Strikingly, lattice dislocations host symmetry-protected gapless states as a consequence of weak indices~\cite{AshvinScrewTI,TeoKaneDefect,ran2010weak,OurDefectPaper}.

Recently, interest has surged in non-Hermitian (nH) topological phases~\cite{KunstRMP,UedaReview}, motivated by their realization in photonic systems~\cite{makris2008,bennet2011,peschel2012,jing2014,feng2014,hodaei2014,peng2014,wu2019,PTreview} and open quantum systems~\cite{naghiloo2019,Minganti2019,jaramillo2020}. The complex-valued spectra of nH systems permit both line- and point-gaps: the former separates the spectrum into two disconnected regions while the latter constitutes a region centered around some reference energy $E$ that contains no eigenstates. Since only point-gapped systems may not be continuously deformable to Hermitian systems without closing the gap~\cite{UedaPaper18,KawabataClassification19}, point-gap topology is intrinsically nH and has a richer classification than its Hermitian counterpart~\cite{KawabataClassification19,HarryPeriodicTable19}. nH topological bands can exhibit distinctive phenomena, including strong sensitivity to boundary conditions via the nH skin effect~\cite{Lee:2016,Torres:2018,YaoZhong18,LonghiSkin19,Jin19,YaoZhongSkin19,Lee:2019,HerviouSVD19,TopoSkin20,zhang2020correspondence,Sato-PRL:2020,Yoshida20,Borgnia20,Kawabata20HOskin,Vecsei21,zhang2021universal}, and topologically stable spectral degeneracies at generic points in the BZ, known as exceptional points (EPs) where eigenstates coalesce~\cite{BerryEP,HeissEP,NoriNH17,XuRings2017,FuNH18,bergholtz2018,moors2019,yang2019semi,kawabata2019semi,Lin19,xue2020,denner2020exceptional,yang2021fermion}. nH phenomena have been experimentally observed in a variety of platforms~\cite{Zhou2018arcs,cerjan2019,corentin2019,Ghatak29561,HofmannReciprocal20,Xiao2020,Helbig20Skin,weidemann2020topological,LiCriticalSkin20,palacios2020,stegmaier2020topological}.

While topological phases of nH crystal defect Hamiltonians have been classified~\cite{NHDefectClass19}, the response of topological nH band structures to crystal defects -- which themselves may induce gap closings and serve as probes for distinguishing distinct Hermitian topological phases~\cite{AshvinScrewTI,TeoKaneDefect,ran2010weak,Vlad2D,VladScrewTI,slager2019,OurDefectPaper} -- remains largely unexplored (see however Refs.~\onlinecite{sun2021geometric,Panigrahi21}). 
Here, we show that lattice dislocations directly probe intrinsic point-gap nH topology via a dislocation non-Hermitian skin effect (DNHSE) in the presence of \emph{periodic boundary conditions in all directions}: introducing a pair of dislocations results in the accumulation of $\mathcal{O}(L)$ eigenmodes localised at one (both) of the dislocations for non-reciprocal (reciprocal) nH systems. We study systems with weak and strong nH topology in two dimensions (2D) and identify bulk invariants predicting the DNHSE. In stark contrast to Hermitian systems, we find that \textit{gapless} nH systems with bulk EPs can exhibit a robust DNHSE. Hence, lattice dislocations provide crisp spectroscopic signatures of intrinsically nH bulk topology. 

\emph{DNHSE from weak topology}--- Consider the $1 \times 1$ Bloch Hamiltonian
\begin{equation} \label{eq: weakwindingHam}
H(\bs{k}) = t_r e^{\mathrm{i} k_x} + t_l e^{-\mathrm{i} k_x} + t_u e^{\mathrm{i} k_y} + t_d e^{-\mathrm{i} k_y},
\end{equation}
with $t_r = t_l^*$, $t_u = t_d^*$ its Hermitian limit. This model is characterized by weak winding number topological invariants:
\begin{equation} \label{eq:1D invariant}
w_{j}(E) = \int_{\mathrm{BZ}}  \frac{d^2\bs{k}}{(2 \pi)^2 \mathrm{i}} \mathrm{Tr}\left\{[H(\bs{k})-E]^{-1} \frac{\partial}{\partial{k_{j}}} [H(\bs{k})-E]\right\},
\end{equation}
where $\mathrm{BZ} = [0,2\pi]^{\times 2}$, and $E$ is a reference energy.
The pair $\bs{w}(E) = [w_{x}(E), w_{y}(E)] \in \mathbb{Z}^{\times 2}$ is quantized for a point-gap at $E$, and indicates weak nH topology: a system with nonzero $\bs{w}$ is adiabatically connected to a disconnected set of 1D Hatano-Nelson chains~\cite{HatanoNelson96,HatanoNelson97,HatanoNelson98}, stacked perpendicular to $\bs{w}(E)$. Importantly, systems with a line-gap connecting to $E$ necessarily exhibit $\bs{w}(E) = \bs{0}$, implying that a nonzero $\bs{w}(E)$ indicates intrinsically nH point-gap topology.

\begin{figure}[t]
\centering
\includegraphics[width=0.533\textwidth]{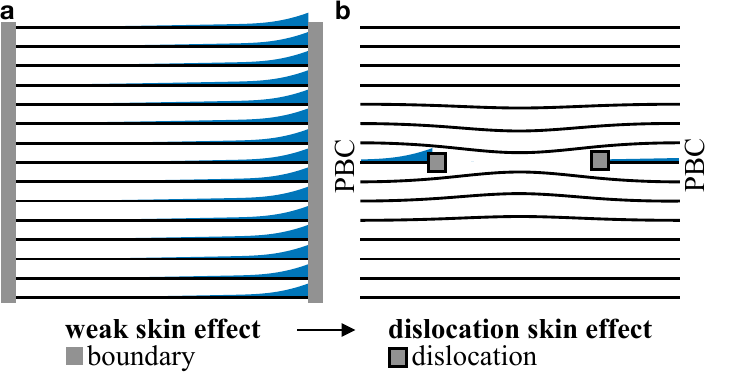}
\caption{Correspondence of weak and dislocation non-Hermitian skin effects. (a)~A 2D system with nontrivial weak winding number can be adiabatically related to a decoupled stack of Hatano-Nelson chains. These exhibit a non-Hermitian skin effect in presence of open boundary conditions. (b)~With periodic boundary conditions (PBC), the same system gives rise to a dislocation non-Hermitian skin effect when a pair of dislocations is introduced into the lattice. The skin effect persists even when couplings along the stacking direction are turned on.}
\label{fig: dislocskin}
\end{figure}

\begin{figure}[t]
\centering
\includegraphics[width=0.48\textwidth]{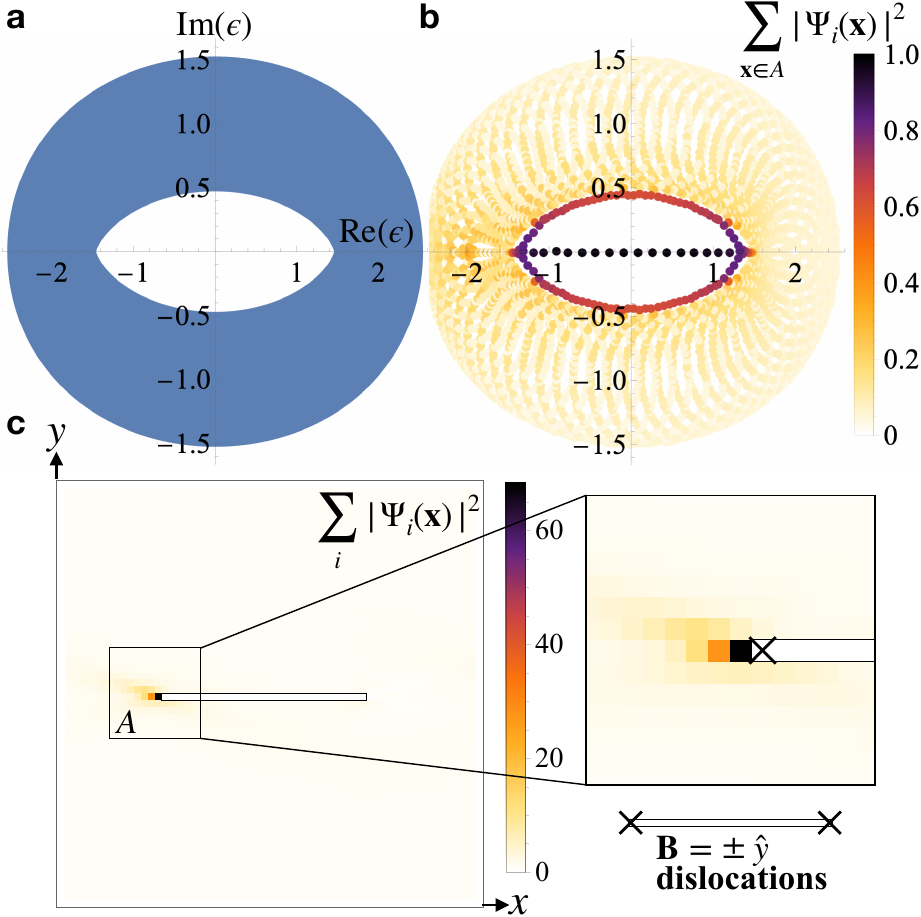}
\caption{Tight-binding model exhibiting weak non-Hermitian topology and a dislocation non-Hermitian skin effect. (a)~PBC spectrum of the Bloch Hamiltonian in Eq.~\eqref{eq: weakwindingHam} for the parameter choice $t_r = 3/2$, $t_l = 1/2$, $t_u = 1/2$, $t_d = 0$. (b)~PBC spectrum on a square geometry of $60 \times 60$ sites in presence of a pair of dislocations separated by $30$ sites. States are colored by their weight in the dislocation region $A$ [indicated in panel c)]. We observe a partial spectral collapse onto the real line, driven by dislocation-localized states. (c)~Local density of states. The accumulation at only one of the two dislocations signals the non-Hermitian skin effect.}
\label{fig: weaknumerics}
\end{figure}

Let us discuss the case of decoupled chains along the $x$-direction ($t_u = t_d = 0$). Adding hoppings along the $y$-direction will not affect our results as long as the point-gap remains open. For $|t_r| > |t_l|$, we find a point-gap at $E = 0$ and $\bs{w}(0) = (1,0)$. Since a nonzero winding number corresponds to a nH skin effect~\cite{zhang2020correspondence,TopoSkin20}, extensively many eigenstates will accumulate on the right boundary of the sample in an open geometry [see Fig.~\ref{fig: dislocskin}~a)]. We next study the effect of introducing (edge) dislocations into the 2D bulk. These 0D crystal defects are characterized by a Burgers vector $\bs{B}$: encircling a dislocation counter-clockwise, $\bs{B}$ equals the end point displacement resulting from sweeping out the same trajectory on a pristine lattice~\footnote{We fix the sense vector~\cite{bollmann2012crystal} of all dislocations to point in the positive $z$ direction.}. In order to preserve PBC, the Burgers vectors of all dislocations in the system must sum to zero: this is essential, because open boundary conditions (OBC) induce a conventional skin effect. Numerically, a single pair of dislocations with Burgers vector $\bs{B} = \hat{y}$ is constructed by removing a line of unit cells at constant $y$ from the crystalline lattice (hereafter called the defect line), and reintroducing hoppings between sites on either side. These hoppings can be such that the defect line becomes locally indistinguishable from the pristine lattice except at the dislocations. For the choice $t_u = t_d = 0$, a pair of $\bs{B} = \hat{y}$ dislocations essentially implements OBC for one chain at a fixed $y$ coordinate, without affecting any remaining chains of the stack [see Fig.~\ref{fig: dislocskin}~b)]. That is, the effective Hamiltonian governing the dislocation response is the 1D Hatano Nelson chain,
\begin{equation} \label{eq: 1DHatanoNelson}
h(k_x) = t_r e^{\mathrm{i} k_x} + t_l e^{-\mathrm{i} k_x},
\end{equation}
\emph{with OBC}. Correspondingly, there is a 1D skin effect, and most eigenstates accumulate at a dislocation (for $|t_r| > |t_l|$, this is the dislocation at the left end of the defect line). Concomitantly, the energy spectrum of the full system -- which previously encircled the point-gap at $E=0$, see Fig.~\ref{fig: weaknumerics}~a) -- undergoes a partial spectral collapse onto the real line, shown in Fig.~\ref{fig: weaknumerics}~b)~\cite{Zhang:2020, Sato-PRL:2020}. The response at only one of the two dislocations [Fig.~\ref{fig: weaknumerics}~c)] characterizes the intrinsically nH topology.

\begin{figure}[t]
\centering
\includegraphics[width=0.48\textwidth]{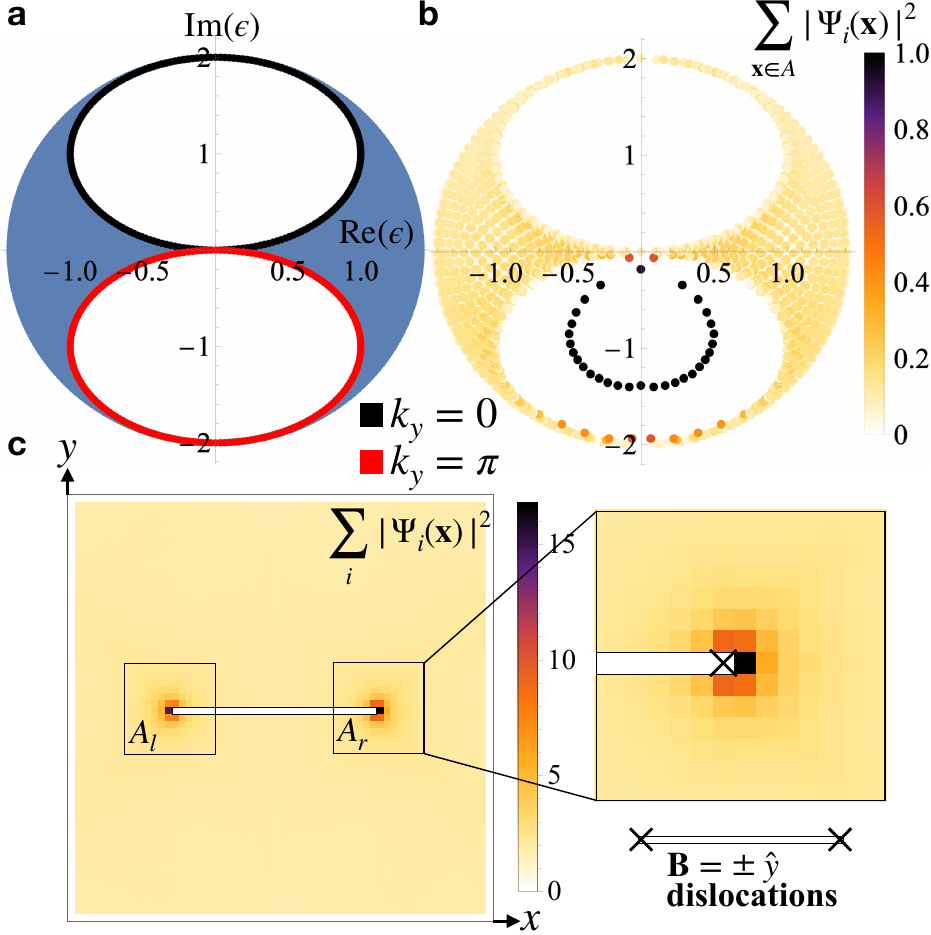}
\caption{Tight-binding model exhibiting strong non-Hermitian topology and a dislocation non-Hermitian skin effect. (a)~Periodic boundary condition (PBC) spectrum of the Bloch Hamiltonian in Eq.~\eqref{eq: strongZ2model} for the parameter choice $t_x = t_y = \gamma = 1$. (b)~PBC spectrum on a square geometry of $60 \times 60$ sites in presence of a pair of dislocations separated by $30$ sites. States are colored by their weight in the dislocation region $A = A_l \cup A_r$ [see panel c)]. Importantly, only the $k_y = \pi$ point-gap [indicated in red in panel a)] participates in the response to dislocations with Burgers vectors $\bs{B} = \pm \hat{y}$. (c)~Local density of states. The accumulation at both of the two dislocations signals the non-Hermitian $\mathbb{Z}_2$ skin effect.}
\label{fig: strongnumerics}
\end{figure}

\emph{DNHSE from strong topology}--- We next investigate the dislocation response of point-gapped strong nH topological phases. Consulting the classification of nH insulators in the 38 Bernard-LeClair classes~\cite{KawabataClassification19}, most symmetry classes in 2D do not have intrinsic point-gap topology~\cite{TopoSkin20}. One exception is symmetry class AII$^\dagger$, characterized by the presence of reciprocity: $[T H(\bs{k})^\mathrm{T} T^\dagger = H(-\bs{k})]$ holds for a unitary operator $T$ $[T T^* = -1]$ and Bloch Hamiltonian $H(\bs{k})$.
Point-gaps in this symmetry class have a $\mathbb{Z}_2$ classification. A Hamiltonian realizing the nontrivial phase is~\cite{TopoSkin20}
\begin{equation} \label{eq: strongZ2model}
H(\bs{k}) = t_x \sin k_x \sigma_x + t_y \sin k_y \sigma_y + \mathrm{i} \gamma (\cos k_x + \cos k_y) \sigma_0,
\end{equation}
where $\sigma_i$, $i = 0,x,y,z$, are the Pauli matrices, $t_x$ and $t_y$ are hopping amplitudes, and $\gamma$ characterizes the strength of non-Hermiticity.
$H(\bs{k})$ is reciprocal for $T = \mathrm{i}\sigma_y$. For nonzero $\gamma$, this model exhibits two point-gaps situated at the complex energies $E^\pm = \pm \mathrm{i} \gamma$ [see Fig.~\ref{fig: strongnumerics}~a)]. Either can be used to evaluate the strong $\mathbb{Z}_2$ topological invariant~\cite{KawabataClassification19}
\begin{equation} \label{eq: 2Dstronginvariant}
(-1)^{\nu(E)} = v_x (E,0) v_x (E,\pi).
\end{equation}
Here, we have defined the $k_y$-resolved $\mathbb{Z}_2$ winding number~\cite{TopoSkin20} in $x$-direction as
\begin{equation}
\begin{aligned}
v_x (E,&k_y) = \mathrm{sgn} \Bigg\{\frac{\mathrm{Pf}[Q(\pi,k_y)]}{\mathrm{Pf}[Q(0,k_y)]}
\\
&\times\exp \bigg[-\frac{1}{2} \int_{k_x = 0}^{k_x = \pi} \mathrm{Tr}\left[Q(\bs{k})^{-1} \frac{\partial}{\partial{k_{x}}} Q(\bs{k})\right]\Bigg\},
\end{aligned}
\end{equation}
where $Q(\bs{k}) = [H(\bs{k})-E]T$, and $\mathrm{Pf}(M)$ is the Pfaffian of the anti-symmetric matrix $M$.
In 1D, a nontrivial $\mathbb{Z}_2$ winding number indicates the presence of a $\mathbb{Z}_2$ skin effect protected by reciprocity~\cite{TopoSkin20}. For a 2D system in symmetry class AII$^\dagger$ to exhibit strong point-gap topology, $v_x (E,0) = - v_x (E,\pi)$ must hold. In particular, for $H(\bs{k})$ in Eq.~\eqref{eq: strongZ2model}, we find $\nu(E^+) = \nu(E^-) = 1$, implying that this model is topological with respect to either point-gap. 

For Hermitian systems, dislocations with Burgers vector $\bs{B}$ probe the topology of BZ submanifolds satisfying $\bs{B} \cdot \bs{k} \mod 2\pi= \pi$. For instance, dislocations bind gapless states in 2D topological insulators only when the 1D BZ line satisfying $\bs{B} \cdot \bs{k} \mod 2\pi= \pi$ carries a nonzero time-reversal polarization~\cite{Zaanen12,z2spinpumpfukane}. Moreover, edge or screw dislocations in 3D insulators bind gapless helical modes iff the 2D BZ plane satisfying $\bs{B} \cdot \bs{k} \mod 2\pi= \pi$ realizes a 2D topological insulator~\cite{AshvinScrewTI}. The correspondence between the topology of BZ submanifolds and dislocation responses was derived in Refs.~\onlinecite{AshvinScrewTI,TeoKaneDefect,ran2010weak}. In the Supplemental Material~\footnote{See Supplemental Material for the $\mathcal{O}(L)$ scaling of the DNHSE, robustness of the DNHSE to large nH perturbations and OBC, and a general proof of the relationship of the DNHSE with bulk point-gap topology.}, we show that it also applies to nH insulators. For example, for the weak-topological nH system in Eq.~\eqref{eq: weakwindingHam} exhibiting a weak winding number $w_x  = 1$ for $|t_r| > |t_l|$, we demonstrated a $\bs{B} = \pm\hat{y}$ DNHSE, which can be attributed to the nontrivial winding number of the 1D BZ line satisfying $k_y = \pi$ [as modeled by the 1D Hamiltonian in Eq.~\eqref{eq: 1DHatanoNelson}]. Interestingly, the strong-topological system in Eq.~\eqref{eq: strongZ2model} exhibits
\begin{equation}
\begin{aligned}
v_x (E^+,0) &= -1, \quad v_x (E^+,\pi) = +1,\\  v_x (E^-,0) &= +1, \quad v_x (E^-,\pi) = -1.
\end{aligned}
\end{equation}
This implies that the $k_y = \pi$ line has trivial (nontrivial) $\mathbb{Z}_2$ winding with respect to the point-gap centered at $E^+$ ($E^-$). Since dislocations with Burgers vector $\bs{B} = \pm \hat{y}$ probe the $k_y = \pi$ line, the point-gap at $E^-$, but not that at $E^+$, contributes a $\mathbb{Z}_2$ DNHSE protected by reciprocity. Indeed, upon introducing a pair of $\bs{B} = \pm \hat{y}$ dislocations into the crystalline lattice, only the point-gap at $E^-$ undergoes a spectral deformation, while the point-gap at $E^+$ remains essentially unaffected [Fig.~\ref{fig: strongnumerics}~b)]. Moreover, extensively many eigenstates accumulate at both dislocations in Fig.~\ref{fig: strongnumerics}~c), as expected from reciprocity. Note that the model in Eq.~\eqref{eq: strongZ2model} -- unlike the system in Eq.~\eqref{eq: weakwindingHam} -- does not exhibit a conventional skin effect with OBC in two directions~\cite{TopoSkin20}. This can be understood intuitively by noting that -- in contrast to the Hatano-Nelson chain -- there is no anomalous charge accumulation in OBC, as edge modes can circulate around the closed 1D boundary of the 2D sample~\cite{TopoSkin20,okuma2020,Kawabata20HOskin,okuma2021}.

\begin{figure*}[t]
\centering
\includegraphics[width=\textwidth]{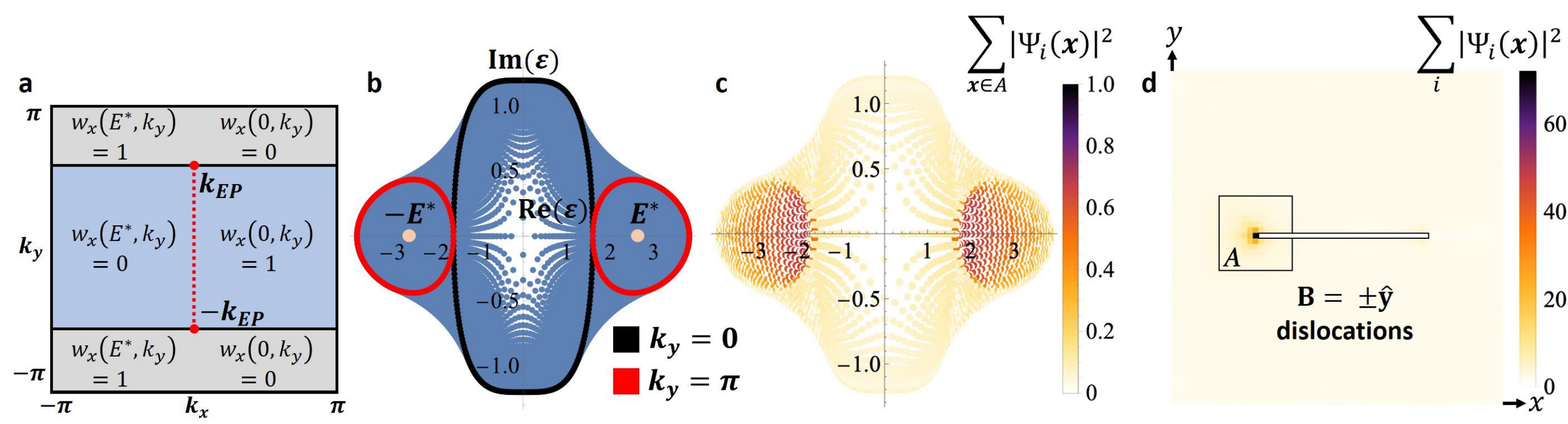}
\caption{Tight-binding model exhibiting stable exceptional points (EPs) and a dislocation non-Hermitian skin effect. (a)~EPs (red dots) are located along $k_x = 0$ for any $\delta \neq 0$. The winding number at $k_y = \pi$ is $1$ ($0$) for the point-gap at $E = E^*$ $(E = 0)$ and vice-versa for $k_y = 0$. (b)~PBC spectrum for the Hamiltonian in Eq.~\eqref{eq:EPHam} for the parameter choice $t_x=t_y=t=1,m=2,\delta = -1.2$, with orange dots denoting $\pm E^*$. (c)~PBC spectrum on a square geometry of $60 \times 60$ sites in presence of a pair of dislocations separated by $30$ sites. States are colored by their weight in the dislocation region $A$ [see panel d)]. Importantly, only the $k_y = \pi$ point-gaps [indicated in red in panel b)] participate in the response to a dislocation with Burgers vector $\bs{B} = \hat{y}$. (d)~Local density of states. The accumulation at only one of the dislocations signals the non-Hermitian skin effect.}
\label{fig:EP}
\end{figure*}

We emphasize that the nH dislocation response is distinct from that of Hermitian insulators with strong topology: in that case, $\bs{B} = \pm \hat{y}$ dislocations only carry bound states if the band inversion occurs within the $k_y = \pi$ line. On the other hand, for the system in Eq.~\eqref{eq: strongZ2model}, both $k_y = 0,\pi$ lines carry nontrivial $\mathbb{Z}_2$ winding numbers, \emph{but with respect to different point-gaps}. Since the DNHSE is a property of the full complex band structure, it is sensitive to both point-gaps. To predict it in general nH systems, we must therefore examine the topology of all point-gaps of the system.

\emph{Dislocation response in gapless systems}--- We study the dislocation response of systems with EPs through:
\begin{equation}
\label{eq:EPHam}
    H(\bs{k}) = \sum_{j=x,y} \left( t_j \sin k_j \sigma_j - t \cos k_j \sigma_z \right)  + i \delta \sigma_x + m \sigma_z
\end{equation}
with anisotropic non-Hermiticity $\delta$ and $t_x,t_y,m,t,\delta \in \mathbb{R}$ (we set $t_x=t_y=1$). This model, which is point-gapped for $\delta \in \left(-|m-2t|,|m+2t| \right)$, is an example of a nH Chern insulator~\cite{yao2018chern,kawabata2018anomalous}. Since Eq.~\eqref{eq:EPHam} respects generalized inversion [$\sigma_x H(\bs{k}) \sigma_x = - H^\dagger(\bs{-k})$] and parity-particle-hole [$\sigma_y H(\bs{k})\sigma_y = H(\bs{-k})$] symmetries, any band crossings necessarily occur at $E=0$~\cite{Okugawa2021}. We fix $m=2t$ and $\delta \in [-2,2]$ such that for $\delta=0$, the point-gap closes and a Dirac cone develops at $\bs{k}=\bs{0}$, while for $\delta \neq 0$, there exists a pair of topologically stable EPs at $\pm \bs{k}_{EP}=(0,\pm \Delta/2)$ with $\Delta=\cos^{-1}\left[ \frac{1}{2} (2-\delta^2) \right]$ [see Fig.~\ref{fig:EP}~a)], and the spectrum remains gapped elsewhere.

These EPs are characterized by a winding number topological invariant ($j \neq j'$)~\cite{kawabata2019semi}:
\begin{equation}
w_j(E,k_{j'})= \int_{k_j \in \mathrm{BZ}} \frac{dk_j}{2\pi \mathrm{i}} \mathrm{Tr}[H(\bs{k})-E]^{-1} \partial_{k_{j}} [H(\bs{k})-E] ,
\end{equation} 
which probes whether the effective 1D Hamiltonian (gapped away from the EPs at $E=0$) with fixed $k_{j'}$ winds along the $j$-cycle and is quantized for a point-gap~\footnote{This is the point-gap for the effective 1D Hamiltonian resulting from treating $k_{j'}$ as a parameter.} at $E$. For the Hamiltonian Eq.~\eqref{eq:EPHam}, $w_y(E,k_x)$ vanishes everywhere while $w_x(E,k_y)$ displays non-trivial behaviour -- since the EPs at $E=0$ signal topological phase transitions in the BZ, $w_x(0,k_y)$ necessarily jumps whenever $k_y$ crosses an EP:
\begin{equation}
w_x(0,k_y) = \begin{cases}
1, & k_y \in (-\Delta/2,\Delta/2)\\
0, & k_y \in (-\pi, -\Delta/2) \cup (\Delta/2,\pi) .
\end{cases}
\end{equation}
Consistent with this, as a function of $k_y$ the eigenvalues of $H$ either trace out a single circle around $E = 0$ or two circles around $\pm E^* \neq 0$ in the complex plane, with transitions at the EPs [see Fig.~\ref{fig:EP}~b)].

Consider a pair of dislocations with Burgers vector $\bs{B} = \pm \hat{y}$, sensitive to the topology of the 1D BZ submanifold specified by $k_y = \pi$. While the trivial $w_x(0,\pi)$ for any $\delta$ suggests the absence of a DNHSE, we identify a non-trivial winding of the $k_y = \pi$ line with respect to the point-gaps at $\pm E^*$: $w_x(\pm E^*, \pi) = 1$. The $k_y = \pi$ line displays a nontrivial (trivial) $\mathbb{Z}$ winding with respect to the point-gap at $\pm E^*$ ($0$) and so we expect that only the point-gaps at $\pm E^*$ contribute a DNHSE for dislocations with $\bs{B} = \pm \hat{y}$. This prediction is vindicated in Fig.~\ref{fig:EP}~c): only eigenstates in the point-gaps at $k_y = \pi$ and $\pm E^*=\pm 3$ become defect-localized. Eigenstates accumulate at only one dislocation [Fig.~\ref{fig:EP}~e)] [for $\delta < 0\,(> 0)$, the skin effect is present at the left (right) dislocation]. No skin effect appears for a pair of $B = \pm \hat{x}$ dislocations, consistent with a trivial $w_y(E,k_x)$.

To verify the topological origin of the DNHSE, we study the total spectral weight near the dislocation core as a function of the distance between $k_y = \pi$ and the EPs at $k_y = \pm\Delta/2$. As shown in the Supplemental Material~\cite{Note3}, the spectral weight is largely independent of this distance when the EPs are well-separated, supporting the fact that only the region around the $k_y = \pi$ line contributes to the DNHSE. 
Finally, we test the stability of the DNHSE by turning on a random nH perturbation (see Supplemental Material~\cite{Note3}). Since EPs are generically stable~\cite{kawabata2019semi}, $w_x(\tilde{E},\pi) = 1$ around some point-gap $\tilde{E} \neq 0$ and the DNHSE persists even for an $O(1)$ perturbation. (We note that, while the conventional skin effect may exist in gapless nH systems~\cite{zhang2021universal}, arguments for its topological origin do not generalize to the DNHSE.)

\emph{Conclusion}---
We have shown that an intrinsically nH DNHSE is present as long as the BZ line satisfying $\bs{B} \cdot \bs{k} \mod 2\pi = \pi$ has a non-trivial ($\mathbb{Z}$ or $\mathbb{Z}_2$) winding around \textit{at least one} point-gap of a nH system. Crucially, for the EP system, the absence of a bulk-gap precludes the usual arguments predicting boundary or defect modes in both Hermitian and nH systems. Nonetheless, we find that dislocations in nH systems display the DNHSE even in the presence of EPs, i.e., $O(L)$ skin modes remain bound to the dislocation core. Our work motivates further study of the topological origin of the dislocation response of EPs, and of the interplay between multiple point-gaps.

\begin{acknowledgments}
FS thanks Nicolas Regnault for helpful discussions, and Benjamin J. Wieder for collaboration on a related topic. AP thanks Biao Lian for discussions on non-Hermitian systems. We were supported by fellowships at the Princeton Center for Theoretical Science.
\end{acknowledgments}


\bibliography{refs}


\end{document}


\title{Supplementary Material for ``Dislocation Non-Hermitian Skin Effect"}

\author{Frank Schindler}
\affiliation{Princeton Center for Theoretical Science, Princeton University, Princeton, NJ 08544, USA}

\author{Abhinav Prem}
\affiliation{Princeton Center for Theoretical Science, Princeton University, Princeton, NJ 08544, USA}

\maketitle


\renewcommand{\bibnumfmt}[1]{[S#1]}
\renewcommand{\theequation}{S\arabic{equation}}
\renewcommand{\thefigure}{S\arabic{figure}}
\renewcommand{\thetable}{S.\Roman{table}}

\setcounter{equation}{0}
\setcounter{figure}{0}
\setcounter{table}{0}

\onecolumngrid


\tableofcontents


\section{Scaling behaviour of the dislocation non-Hermitian skin effect}
\label{sec:scaling}

The skin effect is a well-established signature of topologically non-trivial nH systems~\cite{zhang2020correspondence,TopoSkin20}, wherein $\mathcal{O}(L^{d-1})$ eigenstates localise on the $(d-1)$-dimensional boundary of a $d$-dimensional nH system. For a $d$ dimensional system, if we open boundaries along the $j^{th}$ direction, the skin effect presents at one (both) boundaries for a non-reciprocal (reciprocal) system. Note that $L$ denotes the linear extent of the system. 

\begin{figure}[t]
    \centering
    \includegraphics[width=\textwidth]{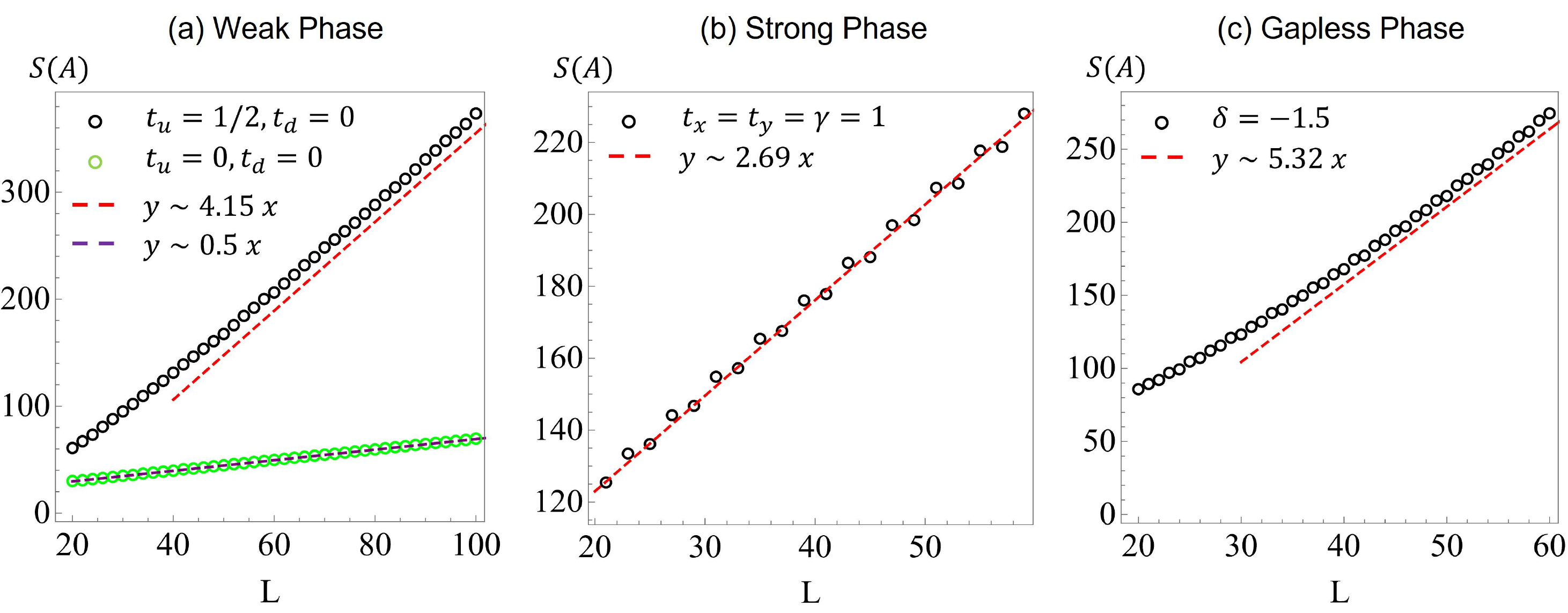}
    \caption{Scaling of the total spectral weight $S(A)$ in a region $A$ localised around the dislocation core(s) for each of the three models considered in the main text. Open dots are results obtained numerically while dashed lines illustrate the linear scaling. For each $L$, we place the system on a square geometry ($L \times L$ sites) with periodic boundary conditions in both $x$ and $y$ directions and introduce a pair of dislocations separated by $\left \lfloor{L/2}\right \rfloor$ and with $\bs{B} = \pm \hat{y}$.}
    \label{fig:scaling}
\end{figure}

Here, we introduce a pair of dislocations with Burgers vector $\bs{B}=\pm\hat{y}$ into each of the three models considered in the main text and explicitly show that the dislocation non-Hermitian skin effect (DNHSE) corresponds to $\mathcal{O}(L)$ eigenstates localising at the dislocation core(s). Specifically, we calculate the total spectral weight of eigenstates $\Psi_i$
\begin{equation}
    S(A) = \sum_{{\bs x} \in A} s({\bs x}), \quad s({\bs x}) = \sum_i |\Psi_i({\bs x})|^2 ,
\end{equation}
over some region $A$ localized near the dislocation core(s) and study it as a function of system size $L$. For each of the three models in the main text, the results are shown in Fig.~\ref{fig:scaling}, with the corresponding region $A$ depicted in Figs.~2~c),~3~c), and~4~d). Note that we keep the size of the region $A$ fixed even as we change the system size. 

\begin{figure}[t]
    \centering
    \includegraphics[width=0.4\textwidth]{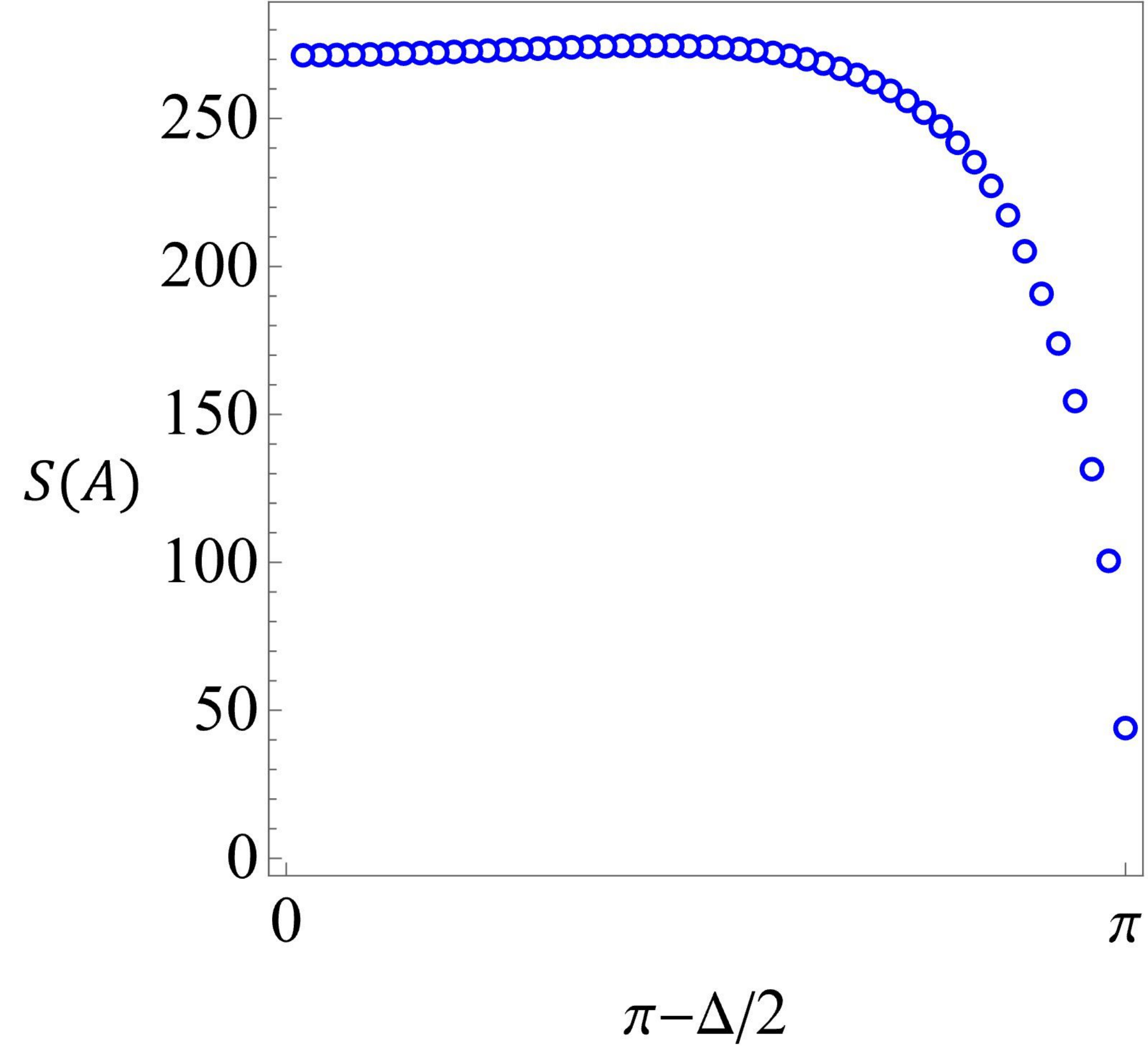}
    \caption{Total spectral weight $S(A)$ as a function of the distance between $k_y = \pi$ and the EP at $k_y = \Delta/2$ for the model in Eq.~(9). We place the system on a square geometry ($60 \times 60$ sites) with periodic boundary conditions in both $x$ and $y$ directions and with a pair of dislocations separated by 30 sites. The area $A$ is shown in Fig.~4~d).}
    \label{fig:EPdist}
\end{figure}

\begin{figure}[t]
\centering
\includegraphics[width=\textwidth]{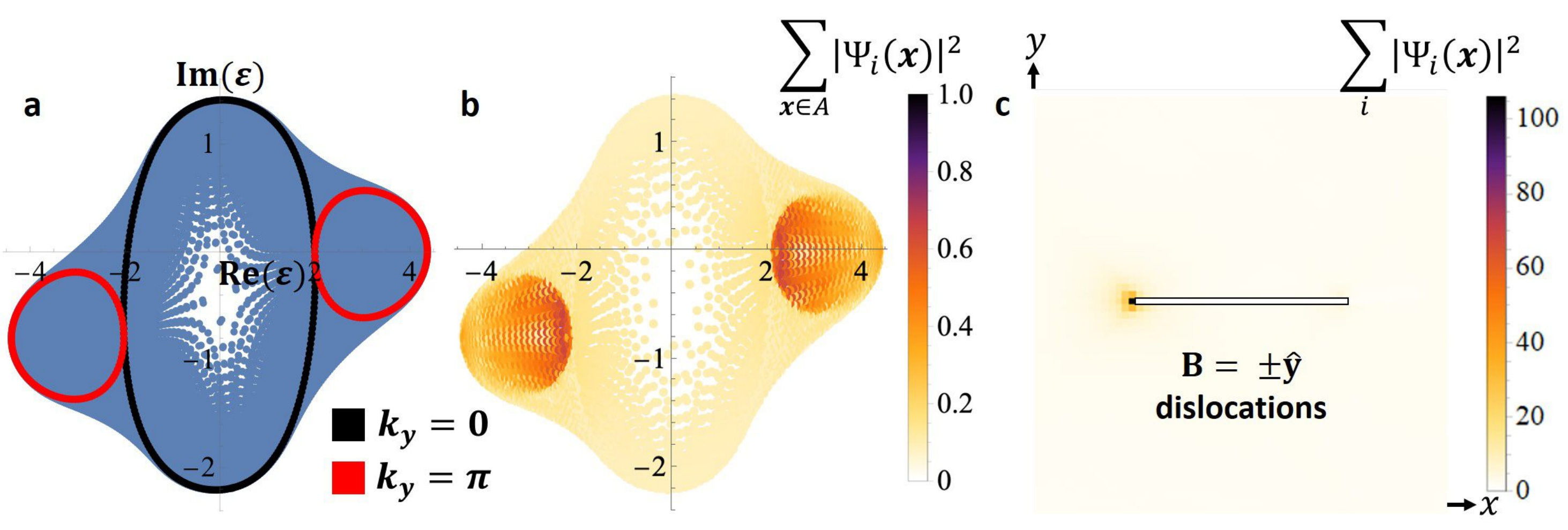}
\caption{Tight-binding model exhibiting stable exceptional points (EPs) and a dislocation non-Hermitian skin effect in the presence of an $\mathcal{O}(1)$ non-Hermitian perturbation [$\delta h$ in Eq.~\eqref{eq:EPHampert}]. (a)~PBC spectrum for the Hamiltonian in Eq.~\eqref{eq:EPHampert} for the parameter choice $t_x=t_y=t=1,m=2,\delta = -1.2$. (b)~PBC spectrum on a square geometry of $60 \times 60$ sites in presence of a pair of dislocations separated by $30$ sites. States are colored by their weight in the dislocation region $A$ [see panel c)]. Importantly, only the $k_y = \pi$ point gaps [indicated in red in panel a)] participate in the response to a dislocation with Burgers vector $\bs{B} = \hat{y}$. (c)~Local density of states. The accumulation at only one of the dislocations signals the non-Hermitian skin effect.}
\label{fig:EPpert}
\end{figure}

As discussed in the main text, for the model with weak nH topology [Eq.~(1)], the DNHSE can be simply understood as a consequence of the conventional skin effect in the one-dimensional Hatano-Nelson model. Specifically, when $t_y = t_d = 0$, Eq.~(1) reduces to a decoupled stack of one-dimensional Hatano-Nelson chains, and introducing a pair of dislocations with $\bs{B} = \pm \hat{y}$ is equivalent to opening the boundary conditions for a single Hatano-Nelson chain at a fixed $y$ coordinate. Thus, since only a single chain contributes to the DNHSE in the decoupled limit, we expect $S(A) \sim L/2$; this is confirmed numerically as shown in Fig.~\ref{fig:scaling}~a). Once we turn on hopping along the $y$ direction, the other one-dimensional chains also contribute to the DNHSE so that $S(A)$ still scales linearly with $L$ albeit with a parameter dependent slope.

Since the model in Eq.~(5) with strong nH topology is reciprocal, eigenstates localise at both of the dislocation cores and we calculate the combined spectral weight from each of the defects [i.e., $A = A_l \cup A_r$ as shown in Fig.~3~c)]. Fig.~\ref{fig:scaling}~b) clearly shows the $\mathcal{O}(L)$ scaling, consistent with the presence of a DNHSE. This scaling persists as long as there is a non-trivial $\mathbb{Z}_2$ winding along the BZ line at $k_y = \pi$. 

Finally, let us consider the model Eq.~(9) with $m=2, t=1$ which hosts bulk EPs for any $\delta \neq 0$. As a consequence of the non-trivial winding $w_x(\pm E^*,\pi) = 1$ (see main text), we find the presence of a DNHSE at one of the dislocation cores in this model. Surprisingly, we find that the total spectral weight localised at the defect shows linear scaling with system size even in the absence of a bulk gap. While for small system sizes we observe deviation from linear scaling as a consequence of finite-size effects, Fig.~\ref{fig:scaling}~c) clearly demonstrates that $S(A) \sim L$ at larger system sizes. To verify that the non-trivial winding of the $k_y = \pi$ BZ submanifold around $E = \pm E^*$ is indeed responsible for the robust dislocation response, we plot the total spectral weight $S(A)$ as a function of the distance $\pi - \Delta/2$ between $k_y = \pi$ and the EPs at $k_y = \pm \Delta/2$, where $\Delta = \cos^{-1}\left[1/2 (2-\delta^2) \right]$. As shown in Fig.~\ref{fig:EPdist}, we see that $S(A)$ is largely independent of this distance when the two EPs are not too close to $k_y=0$ (in this limit, where $\delta = 0$, the DNHSE must vanish because the system becomes Hermitian); this supports our claim that the non-trivial topological winding of the $k_y = \pi$ BZ submanifold is responsible for the DNHSE even in the absence of a bulk-gap.


\section{Stability of the dislocation non-Hermitian skin effect to large non-Hermitian perturbations}
\label{sec:stability}

Here, we demonstrate the robustness of the DNHSE to generic non-Hermitian perturbations in the model of Eq.~(9) with bulk EPs. In particular, we consider the following perturbed Hamiltonian
\begin{equation}
\label{eq:EPHampert}
    H_{\mathrm{pert}}(\bs{k}) = H(\bs{k}) + \delta h, 
\end{equation}
where $H(\bs{k})$ is given by Eq.~(9) with $t_x = t_y = t = 1, m=2$ and $\delta h$ is a random $2\times 2$ nH matrix. As discussed in the main text, the unperturbed Hamiltonian exhibits topologically non-trivial EPs, which do not rely upon any symmetry/fine tuning. That is, EPs are topologically stable band degeneracies which are robust against any continuous changes in the Hamiltonian as long as they do not cause the system to undergo a phase transition where EPs are created or annihilated~\cite{kawabata2019semi}. Thus, we expect that a generic nH perturbation $\delta h$ will only move the location of the EPs in the BZ but will not cause the pair to annihilate.

Concurrently, the point-gap for the $k_y = \pi$ line will remain open for generic perturbations since the system is unlikely to undergo a phase transition in the 2D BZ. We hence expect that adding a nH perturbation will simply move the location of the EPs and of the energies around which the the $k_y = \pi$ BZ line winds, but will not change the topological winding invariant of the $k_y = \pi$ line. As a consequence, the DNHSE is robust to generic (non fine-tuned) nH perturbations. Note crucially that, unlike in Hermitian systems, we do not impose any requirement on the strength of these perturbations or place any symmetry constraints: the DNHSE in gapless nH systems hence represents a topological phenomenon with no Hermitian counterpart. As shown in Fig.~\ref{fig:EPpert}, we find that the DNHSE is indeed robust to $\mathcal{O}(1)$ (compared to the unperturbed part of the Hamiltonian) nH perturbations.


\section{Dislocation non-Hermitian skin effect under open boundary conditions }
\label{sec:OBC}

For completeness, here we consider the DNHSE for each of the three models considered in the main text in the presence of open boundary conditions (OBC). 

\begin{figure}[t]
    \centering
    \includegraphics[width=0.4\textwidth]{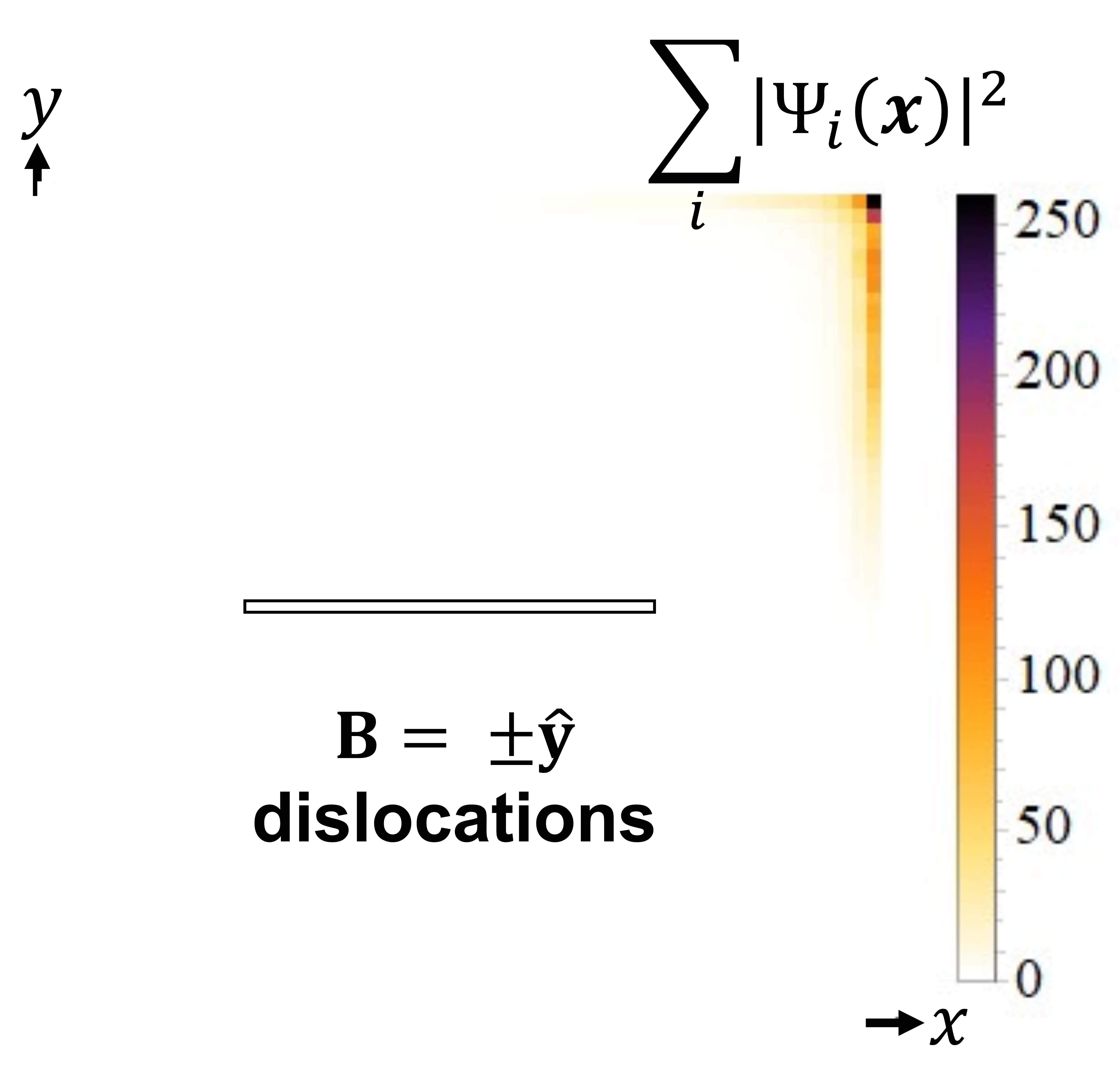}
    \caption{Local density of states for the Hamiltonian in Eq.~\eqref{eq:weakRS} on a square geometry of $60 \times 60$ sites in presence of a pair of dislocations separated by $30$ sites and with OBC. The parameters are the same as those in Fig.~(2) in the main text: $t_r = 3/2, t_l = t_u = 1/2, t_d = 0$.}
    \label{fig:OBCWeak}
\end{figure}

\begin{itemize}

\item The real space Hamiltonian for the weak topological phase, corresponding to Eq.~(1) in the main text, is 
\begin{equation}
\label{eq:weakRS}
    H = \sum_{\bm{r}} \left[ t_l c_{\bm{r}}^\dagger c_{\bm{r} + \bm{\hat{x}}} + t_r c_{\bm{r}+\bm{\hat{x}}}^\dagger c_{\bm{r}} + t_d c_{\bm{r}}^\dagger c_{\bm{r} + \bm{\hat{y}}} + t_u c_{\bm{r} + \bm{\hat{y}}}^\dagger c_{\bm{r}} \right].
\end{equation}
We study this system on a square geometry with OBC in both directions and in the presence of a pair of dislocations. As is clear from Fig.~\ref{fig:OBCWeak}, we see that the DNHSE is greatly suppressed. Indeed, this is exactly what one expects since the model Eq.~\eqref{eq:weakRS} exhibits the conventional skin effect under OBCs and we see clearly that modes are localized around the right boundary. Note that the localization is asymmetric along the $y$-axis since $t_u \neq t_d$ and the modes would be symmetrically localized along the right boundary if $t_u = t_d$.

\item The real space Hamiltonian for the strong topological phase, corresponding to Eq.~(5) in the main text, is 
\begin{equation}
\label{eq:strongRS}
\begin{aligned}
    H = \sum_{\bm{r}} \bigg[&\frac{t_x}{2 i} \left(c_{\bm{r}+\bm{\hat{x}}}^\dagger \sigma_x c_{\bm{r}} - c_{\bm{r}}^\dagger \sigma_x c_{\bm{r}+\bm{\hat{x}}} \right) + \frac{t_y}{2 i} \left(c_{\bm{r}+\bm{\hat{y}}}^\dagger \sigma_y c_{\bm{r}} - c_{\bm{r}}^\dagger \sigma_y c_{\bm{r}+\bm{\hat{y}}} \right) \\
    &+ \frac{i \gamma}{2} \left(c_{\bm{r} + \bm{\hat{x}}}^\dagger c_{\bm{r}} + c_{\bm{r}}^\dagger c_{\bm{r} + \bm{\hat{x}}}  \right) + \frac{i \gamma}{2} \left(c_{\bm{r} + \bm{\hat{y}}}^\dagger c_{\bm{r}} + c_{\bm{r}}^\dagger c_{\bm{r} + \bm{\hat{y}}}  \right) \bigg].
\end{aligned}
\end{equation}
For this model, which exhibits strong nH topology, we see from Fig.~\ref{fig:OBCStrong} that the DNHSE persists and modes remain localized in the vicinity of the dislocation cores. However, we also observe additional modes that are localized at the corners of the sample -- since this model is known not to exhibit the conventional skin effect for full OBC~\cite{okuma2020}, these corner modes are spurious and, unlike the DNHSE, not topologically enforced.

\begin{figure}[t]
    \centering
    \includegraphics[width=0.4\textwidth]{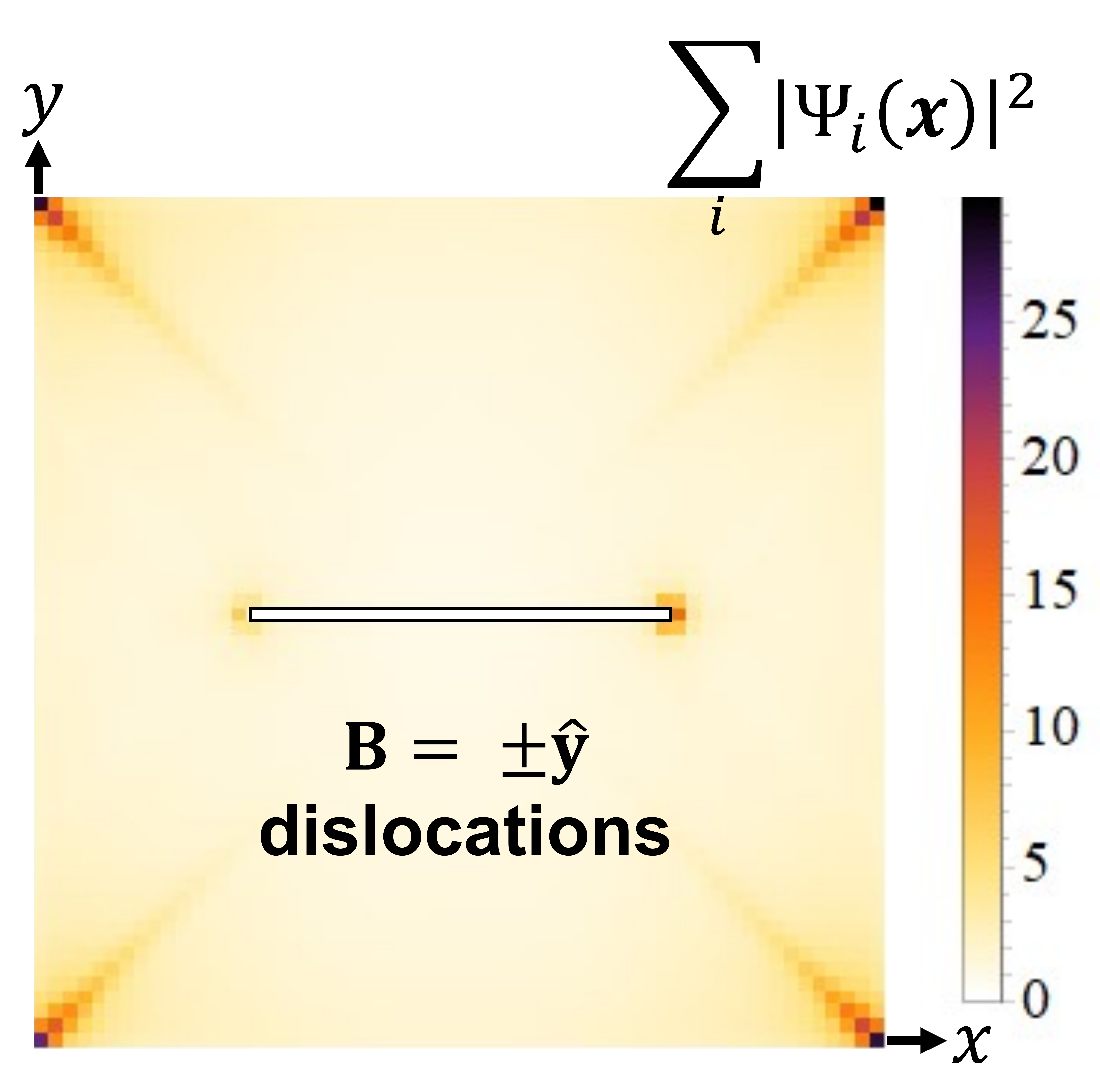}
    \caption{Local density of states for the Hamiltonian in Eq.~\eqref{eq:strongRS} on a square geometry of $60 \times 60$ sites in presence of a pair of dislocations separated by $30$ sites and with OBC. The parameters are the same as those in Fig.~(3) in the main text: $t_x = t_y = \gamma = 1$.}
    \label{fig:OBCStrong}
\end{figure}

\item The real space Hamiltonian for the exceptional point Hamiltonian, corresponding to Eq.~(9) in the main text, is
\begin{equation}
    \label{eq:EPRS}
    H = \sum_{\bm{r}} \left[ c_{\bm{r}}^\dagger \left(m \sigma_z + i \delta \sigma_x \right)c_{\bm{r}} \right] +  \sum_{\bm{r}} \sum_{j=x,y} \left[ c_{\bm{r}}^\dagger \left(\frac{i}{2}t_j \sigma_j -\frac{1}{2}t \sigma_z \right) c_{\bm{r}+ \bm{\hat{j}}}+ \text{h.c.} \right]
\end{equation}

\begin{figure}[t]
    \centering
    \includegraphics[width=0.4\textwidth]{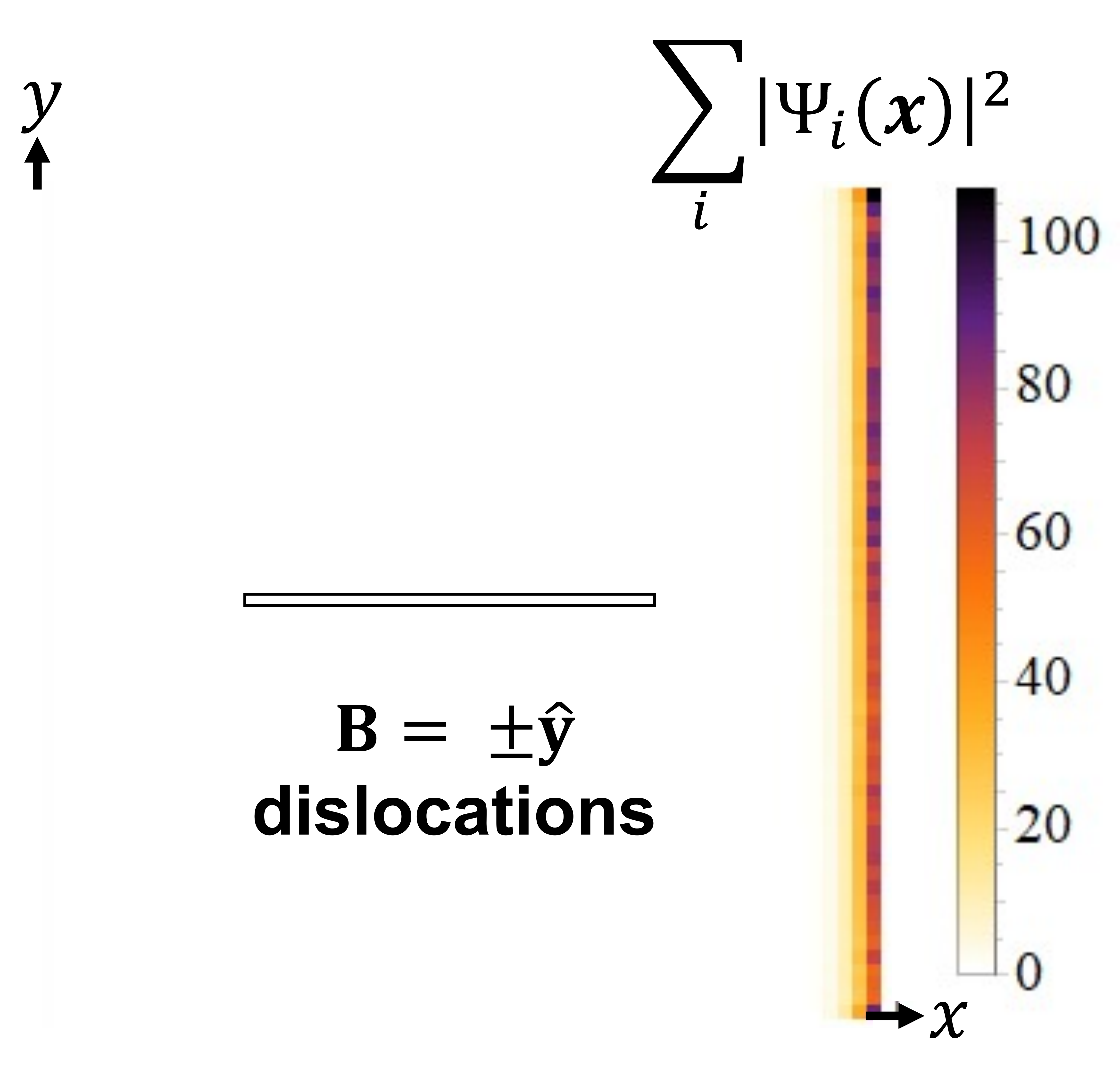}
    \caption{Local density of states for the Hamiltonian in Eq.~\eqref{eq:EPRS} on a square geometry of $60 \times 60$ sites in presence of a pair of dislocations separated by $30$ sites and with OBC. The parameters are the same as those in Fig.~(4) in the main text: $m=2, t_x = t_y = t = 1, \delta=-1.2$.}
    \label{fig:OBCEP}
\end{figure}
As discussed in Ref.~\cite{zhang2021universal}, nH systems with EPs can exhibit the conventional skin effect. This is also the case for the model Eq.~\eqref{eq:EPRS} when placed on a sample with OBC in both directions. As a consequence, we see in Fig.~\ref{fig:OBCEP} that the DNHSE is greatly suppressed due to the competition with the conventional skin effect. 

\end{itemize}


\section{Proof that nontrivial point-gap topology of Brillouin zone submanifolds effects a dislocation non-Hermitian skin effect}
\label{sec:proof}

We here derive necessary and sufficient conditions for the DNHSE in presence of a nH point gap. We focus on 2D nH insulators in two simple symmetry classes: (1) no symmetry and (2) reciprocity. The conditions derived here were exemplified by the models with weak and strong topology described in the main text, but hold more generally in that they can be used to predict the DNHSE for arbitrary nH band structures in the same symmetry class.
Specifically, we prove that the $\mathbb{Z}$ ($\mathbb{Z}_2$) nH winding number of 1D BZ loops in 2D non-reciprocal (reciprocal) nH insulators determines the presence or absence of a $\mathbb{Z}$ ($\mathbb{Z}_2$) DNHSE, via the if-and-only-if relations
\begin{equation} \label{eq: nonhermcorrespondence}
\begin{rcases}
  \nu^{\mathbb{Z}}_{\gamma(\bs{B})}(E) \neq 0 \quad &\leftrightarrow \quad \mathbb{Z} \, \text{DNHSE} \\
  \nu^{\mathbb{Z}_2}_{\gamma(\bs{B})}(E) = 1 \quad &\leftrightarrow \quad \mathbb{Z}_2 \, \text{DNHSE}
\end{rcases} \text{occurs for dislocations with Burgers vector } \bs{B}.
\end{equation}
Here, the 1D BZ loop $\gamma(\bs{B})$ in the Brillouin zone BZ is determined by the condition 
\begin{equation}
\gamma(\bs{B}) = \{\bs{k}\in \mathrm{BZ}|\bs{B} \cdot \bs{k} \mod 2\pi = \pi \}, 
\end{equation}
Moreover, the $\mathbb{Z}$ and $\mathbb{Z}_2$ nH winding numbers with respect to a point gap at complex energy $E$ are defined as
\begin{equation}
    \nu^{\mathbb{Z}}_{\gamma(\bs{B})}(E) = \oint_{\gamma(\bs{B})} \frac{d\bs{k}}{2\pi \mathrm{i}} \cdot \bs{\nabla_k} \log \det \left[H(\bs{k})-E\right] \quad \in \mathbb{Z},
\end{equation}
and
\begin{equation}
    (-1)^{\nu^{\mathbb{Z}_2}_{\gamma(\bs{B})}(E)} = \mathrm{sgn} \Bigg\{\frac{\mathrm{Pf}[Q(\bs{k}_f)]}{\mathrm{Pf}[Q(\bs{k}_i)]} \exp \bigg[-\frac{1}{2} \int_{\bs{k}_i}^{\bs{k}_f} d\bs{k} \cdot \mathrm{Tr}\left[Q(\bs{k})^{-1} \bs{\nabla_k} Q(\bs{k})\right]\Bigg\},
\end{equation}
respectively. In the last expression, we have used $Q(\bs{k}) = [H(\bs{k})-E]T$, where $T$ is the reciprocity operator (see below or main text), $\mathrm{Pf}(M)$ is the Pfaffian of the anti-symmetric matrix $M$, and we have introduced a reciprocal decomposition of the BZ loop $\gamma(\bs{B}) = \{\bs{k}_i \rightarrow \bs{k}_f\} \cup \{-\bs{k}_f \rightarrow -\bs{k}_i\}$ where $\bs{k}_i = -\bs{k}_i+\bs{G}$ and $\bs{k}_f = -\bs{k}_f+\bs{G}$ must hold for a 2D reciprocal lattice vector $\bs{G}$. For consistency, we note that $w_x(E,k_y)$ as defined in Eq.~(11) of the main text satisfies $w_x(E,\pi) = \nu^{\mathbb{Z}}_{\gamma(\hat{y})}(E)$, where the Burgers vector $\hat{y}$ is a unit vector in $y$-direction. Moreover, $v_x(E,k_y)$ as defined in Eq.~(7) of the main text satisfies $v_x(E,\pi)=(-1)^{\nu^{\mathbb{Z}_2}_{\gamma(\hat{y})}(E)}$.

In order to prove Eq.~\eqref{eq: nonhermcorrespondence}, we begin by recalling that a nH Bloch Hamiltonian $H(\bs{k})$ naturally gives rise to an extended Hermitian Hamiltonian $\tilde{H}(\bs{k})$. Given a point gap at complex energy $E$, such an extended Hamiltonian is defined via
\begin{equation} \label{eq: extendedH}
    \tilde{H}(\bs{k}) = \begin{pmatrix}
    0 & H(\bs{k})-E \\
    H(\bs{k})^\dagger -E^* & 0
    \end{pmatrix}.
\end{equation}
Irrespective of $E$, the extended Hamiltonian automatically satisfies a chiral (anti-)symmetry
\begin{equation} \label{eq: chiralsym}
C\tilde{H}(\bs{k}) C^\dagger = - \tilde{H}(\bs{k}), \quad C= \begin{pmatrix} \mathbb{1} & 0 \\ 0 & - \mathbb{1} \end{pmatrix},
\end{equation}
which implies that its eigenvalues are either exactly zero or must come in pairs $(\epsilon,-\epsilon)$. Hence, when $H(\bs{k})$ has no special symmetry itself, the corresponding $\tilde{H}(\bs{k})$ lies in Altland-Zirnbauer symmetry class AIII. On the other hand, if $H(\bs{k})$ is reciprocal so that
\begin{equation} \label{eq:reciprocity}
T H(\bs{k})^{\mathrm{T}} T^\dagger = H(-\bs{k}),
\end{equation}
where $T$ is a unitary operator that satisfies $T T^* = -1$, then the extended Hermitian Hamiltonian has a conventional time-reversal symmetry
\begin{equation}
\tilde{T}\tilde{H}(\bs{k}) \tilde{T}^\dagger = \tilde{H}(-\bs{k}), \quad \tilde{T} = \begin{pmatrix} 0 & T \\ T & 0 \end{pmatrix} \mathit{K}.
\end{equation}
Here, $\mathit{K}$ denotes complex conjugation, so that $\tilde{T}$ is anti-unitary as required for time-reversal. Since $\tilde{T}$ and $C$ anti-commute, any $\tilde{H}(\bs{k})$ associated with a reciprocal nH Hamiltonian lies in Altland-Zirnbauer symmetry class DIII.

Importantly, because the mapping in Eq.~\eqref{eq: extendedH} is on-site in real space, inserting a pair of dislocations into a nH insulator described by $H(\bs{k})$ translates into inserting a pair of dislocations into a Hermitian insulator described by $\tilde{H}(\bs{k})$. The dislocation response of the nH Hamiltonian $H(\bs{k})$ can therefore be induced from two ingredients: (1) the response of the Hermitian Hamiltonian $\tilde{H}(\bs{k})$ and (2) a mapping between the response of $\tilde{H}(\bs{k})$ and the response of $H(\bs{k})$. We discuss these two ingredients in the following.

(1) \emph{Dislocation response of $\tilde{H}(\bs{k})$}. We observe that the invariants $\nu^{\mathbb{Z}}_{\gamma(\bs{B})}(0)$ and $\nu^{\mathbb{Z}_2}_{\gamma(\bs{B})}(0)$ are simply the 1D invariants of Hermitian topological insulators in Altland-Zirnbauer symmetry classes AIII and DIII, respectively: When we identify $\gamma(\bs{B})$ as the Brillouin zone of a 1D Hermitian wire with chiral symmetry, then $\nu^{\mathbb{Z}}_{\gamma(\bs{B})}(0)$ counts the number of protected zero-energy end states, and $\nu^{\mathbb{Z}_2}_{\gamma(\bs{B})}(0)$ counts the parity of gapless Kramers pair end states in presence of time-reversal symmetry. (Here, we have set the reference energy to $E=0$, it is only well-defined relative to a given nH Hamiltonian.) Refs.~\onlinecite{TeoKaneDefect,ran2010weak} related these invariants to the presence of dislocation-localized midgap states in 2D Hermitian topological insulators in the Altland-Zirnbauer symmetry classes AIII and DIII. Specifically, the following correspondence holds:
\begin{equation} \label{eq: hermcorrespondenceran}
\begin{rcases}
  \nu^{\mathbb{Z}}_{\gamma(\bs{B})}(0) \neq 0 \quad &\leftrightarrow \quad \nu^{\mathbb{Z}}_{\gamma(\bs{B})}(0) \, \text{zero-modes} \\
    \nu^{\mathbb{Z}_2}_{\gamma(\bs{B})}(0) = 1 \quad &\leftrightarrow \quad \text{two Kramers-paired zero-modes}
\end{rcases} \text{are bound to dislocations with Burgers vector } \bs{B}.
\end{equation}

(2) \emph{Relationship between $\tilde{H}(\bs{k})$ and $H(\bs{k})$}. To connect the dislocation response of the extended Hermitian Hamiltonian $\tilde{H}(\bs{k})$ with that of the corresponding nH Hamiltonian $H(\bs{k})$, we first note that $E$ can be chosen freely in Eq.~\eqref{eq: extendedH}, as long as it lies in a point gap of $H(\bs{k})$ -- this requirement translates to a finite energy gap of $\tilde{H}(\bs{k})$. Hence, since $\nu^{\mathbb{Z}}_{\gamma(\bs{B})}(E)$ and $\nu^{\mathbb{Z}_2}_{\gamma(\bs{B})}(E)$ are independent of the reference energy $E$ as long as it lies in the point gap, the dislocation response of $\tilde{H}(\bs{k})$ is also independent of $E$. We can now employ the results of Refs.~\onlinecite{TopoSkin20,okuma2020,okuma2021}, where it was shown that the presence of \emph{exact} zero-energy states (Kramers pairs) in the spectrum of the extended Hermitian Hamiltonian for all $E$ inside the point gap is equivalent to a $\mathbb{Z}$ ($\mathbb{Z}_2$) skin effect of the non-Hermitian Hamiltonian. Here, by \emph{exact}, we mean that these zero-modes must be eigenstates of the chiral symmetry in Eq.~\eqref{eq: chiralsym}, as is the case for the protected 0D zero-modes implied in Eq.~\eqref{eq: hermcorrespondenceran}. Hence, the conditions in Eq.~\eqref{eq: hermcorrespondenceran} imply a skin effect in the nH model. Since the system with dislocation is indistinguishable from the pristine crystal described by $H(\bs{k})$ except at the dislocations themselves, this skin effect can only locate at the dislocations cores. Correspondingly, we deduce that $H(\bs{k})$, when put on a lattice hosting a pair of dislocations, exhibits a $\mathbb{Z}$ ($\mathbb{Z}_2$) DNHSE as long as the first (second) condition in Eq.~\eqref{eq: hermcorrespondenceran} is fulfilled. This concludes our proof of Eq.~\eqref{eq: nonhermcorrespondence}.



\bibliography{refs}
